\documentclass[11pt]{article}
\usepackage{graphicx}
 \usepackage{setspace}
\newcommand{\pdfig}[2]{
	\begin{figure}
	\leavevmode	
	\centerline{\includegraphics[width=\columnwidth]{#2.eps}}
	\caption{#1}\label{fig:#2}
	\end{figure}
	\arabic{figure}}

\newcommand{\GS}{{\em GoogleScholar}\ }
\newcommand{\ISI}{{\em ISI}\ }

\begin{document}

\title{Scientific impact quantity and quality: \\
Analysis of two  sources of bibliographic data}

\author{Richard K. Belew \\
Cognitive Science Dept. \\
Univ. California -- San Diego\\
La Jolla CA 92093-0515 USA}

\date{10 April 2005}

\maketitle

\begin{center}{\bf arXiv\#: CoRR/0504036} \end{center}


\begin{quotation}
\begin{center}{\bf Abstract} \end{center}

Attempts to understand the consequence of any individual scientist's
activity within the long-term trajectory of science is one of the most
difficult questions within the philosophy of science.  Because
scientific publications play such as central role in the modern
enterprise of science, bibliometric techniques which measure the
``impact'' of an individual publication as a function of the number of
citations it receives from subsequent authors have provided some of
the most useful empirical data on this question.  Until recently,
Thompson/ISI has provided the only source of large-scale ``inverted''
bibliographic data of the sort required for impact analysis.  In the
end  of 2004, Google introduced a new service, GoogleScholar,
making much of this same data available.  Here we analyze 203
publications, collectively cited by more than 4000 other publications.
We show surprisingly good agreement between data citation counts
provided by the two services.  Data quality across the systems is
analyzed, and potentially useful complementarities between are
considered.  The additional robustness offered by multiple sources of
such data promises to increase the utility of these measurements as
open citation protocols and open access increase their impact on
electronic scientific publication practices.

\end{quotation}

\pagebreak

\section{Background}

Bibliometric analysis of scientific publications goes back to at least
the 1970s \cite{REF621,Price86,REF620}; similar analysis of judicial
opinions has been done by Shepards/LexisNexis for more than a hundred
years.  The Institute for Scientific Information has made an
industry of providing citation data to libraries since the mid-1960s;
the products are currently available as part of Thomson/ISI (\ISI).
\ISI reports that they currently index 16,000 journals, books and
proceedings \cite{garfield98}.  While far from exhaustive (ISI estimates that of the
2000 new journals reviewed annually, only 10\% are selected), the
service cites ``Bradford's Law'' that a relatively small number of
sources capture the bulk of significant scientific results.  All
articles appearing in selected publications have their bibliographies
manually transcribed, and ``inverted bibliographies'' pointing from a
(earlier) cited work to all (subsequent) citing publications is
generated to support users' searches.  Critically, the translation of these
bibliographies into distinct records involves a great deal of {\em
  manual} effort.

May  has reported extensive analyses of British
scientific activity in comparison with other countries, primarily
based on \ISI's data \cite{may97b,may97a}.  ``The database has many shortcomings and
biases, but overall it gives a wide coverage of most fields.''  \cite[p. 793]{may97a}
His critique of shortcomings in this data is useful:
\begin{quotation}
   Some problems have to do with the compilation of the database. It
   includes citations of books and chapters in edited books, but it
   does not include the citations in such publications.  Other
   publications, such as government and other agency reports and
   working papers, are essentially omitted.  It does not cover all
   significant scientific journals....  Papers that describe technical
   methods may attract thousands of reflexive citations, while
   path-breaking papers may be cited only slightly for many years.
   Review articles can mask the primary papers they review. Citation
   patterns vary among fields....  Spectacular scientific errors may
   attract many citations....  Self-citation (which accounts for at
   least 10\% of all citations) may bias some of the results.  \cite[Footnote 3]{may97a}
\end{quotation}
Some of these issues (e.g., having to do with the sources being
compiled) can be expected to altered by new forms of electronic
scientific publication, but others (e.g., self-citation) are likely to
be more intrinsic to scientific authoring processes.  It is for this
reason that Google's recent announcement of their Scholar.Google(beta)
(\GS) service is welcome, as a second, independent source of similar
data.

While specifics concerning Google's operation are difficult to come
by, it is reasonable to assuem that the process relies on more {\em
  automatic}, algorithmic procedures than those used by \ISI.  Linkage
structure among Web pages  is analogous in important ways to
scientific publication \cite{REF1162,Lawrence98}.  These links are
captured by Web crawling algorithms as both ``citing'' pages (i.e.,
Web pages with HTML anchors pointing to other Web pages) and
``cited'' pages are visited, a feature exploited by Google's original
``PageRank'' retrieval algorithm \cite{Page98}.  \GS attempts to bring
similar analyses to academic publication, despite the fact that these
source documents are often much less accessible.

\section{Methods}

Given an author's name\footnote{Translation of an author's name into
  search query string(s) can be ambiguous.  In these experiments both
  first letter, and first letter with the middle initial together with
  full last name was used as the author's name.}, both \ISI and \GS
provide search facilities that return a list of publications
putatively authored by this individual, together with the number of
times each of these publications has been cited by other publications
discovered by the service.  Six academics were selected at random and
used as ``probe'' queries with both systems. \footnote{These academics
  were all drawn from a single, particularly interdisciplinary
  academic department.}  Complete bibliographies of all publications
by these authors were manually reconciled against 203 references to
these publications returned by one or both systems, and then analyzed
in detail.  Cumulatively, \ISI discovered 4741 such references, \GS
found 4045.

Because standards and format of bibliographic citations vary widely
across different publications, the process of reconciling citation
strings from different papers to the same target publication is
problematic, whether via \ISI's manual process or Google's automatic
one.  It is common, therefore, to find the same publication has been
treated as more than one record.\footnote{The alternative type of
  error, where citations to multiple, distinct publications are
  confounded as part of the citation record of a single entry, is 
  more difficult to identify}

For example, manual inspection reveals that a single publication in
the ``Proceedings of the 12th Annual Conference of ACM's Special
Interest Group in Information Retrieval (SIGIR)" is listed as twelve
separate records by \ISI; these are shown in Table 1.  While most
citations to this target publication have been conveniently collected
with respect to two of these records, such noisy data makes impact
analysis difficult.  In these experiments, a publication's ``impact''
is defined as the number of citations found to any of the variations
resolved to the published work, i.e., the sum is taken across all
records (manually) identified as referencing the same publication.

\begin{table}
\begin{center}
\begin{tabular}{|cl|c|}  \hline
{\bf PubYear} & {\bf CiteString} & {\bf NCitations} \\ \hline
1989	& 12 ANN INT ACM SIGIR    	& 1 \\
1989	& 12 ANN INT C RES DEV  	&   1 \\
1989	& 12TH P ANN INT ACM S   11	& 14 \\
1989	& 12TH P INT C RES DEV    	& 1 \\
1989	& ACM SIGIR INT C RES    	& 1 \\
1988	& JUN P ACM SIGIR 88 G   11	& 1 \\
1989	& P 11 INT ACM SIGIR C    	& 1 \\
1989	& P 12 ANN INT ACM SIG    	& 2 \\
1989	& P 12 ANN INT ACM SIG   11	& 16 \\
1989	& SIGIR 89   11	& 2 \\
1989	& SIGIR FORUM 23 11	& 1 \\
1990	& SIGOIS B 11 48	& 1 \\ \hline
\end{tabular}
\caption{Citation variations for same publication}
\end{center}
\end{table}

\section{Results}

Figure \pdfig{Redundant citation noise}{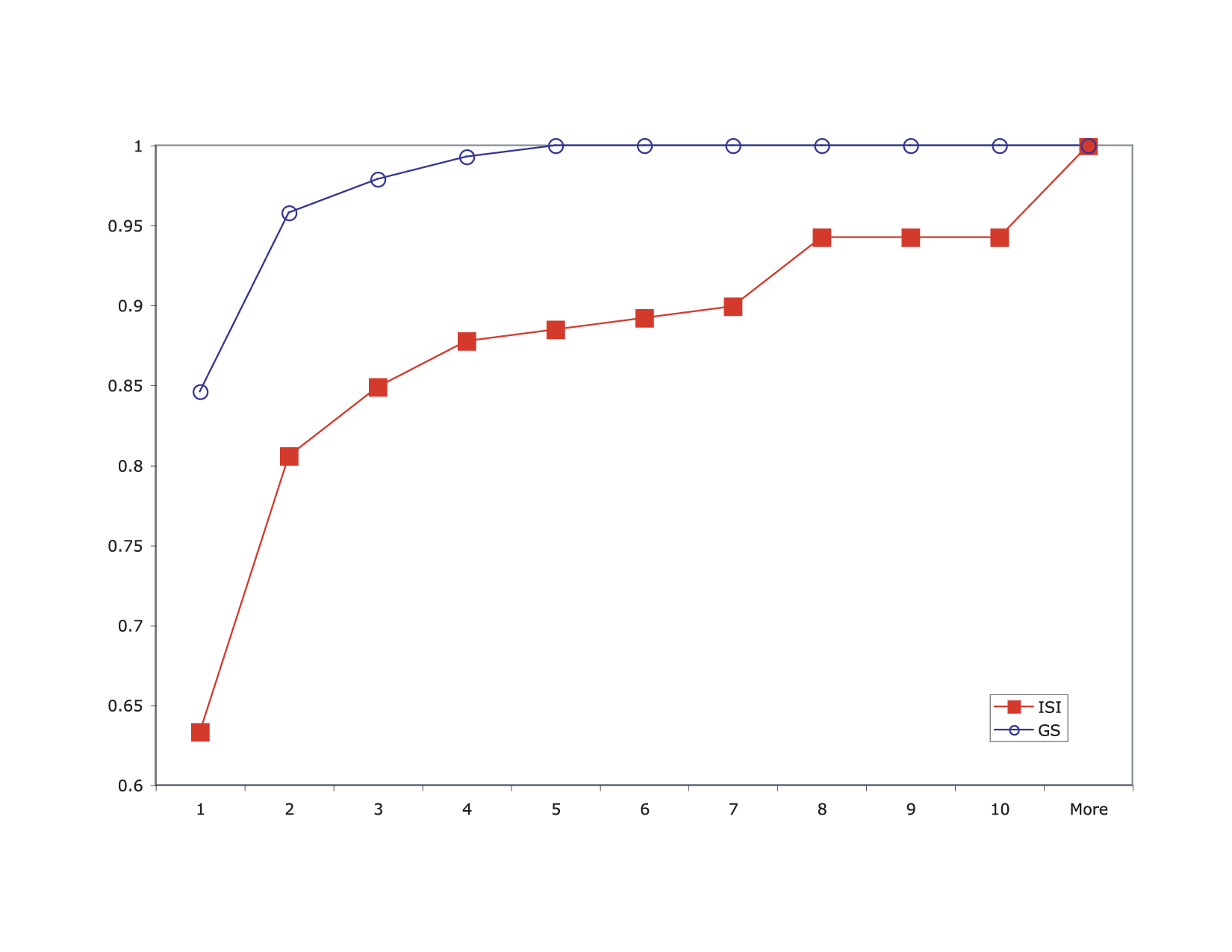} shows how well both
systems aggregate individual citations that in fact to refer to the
same published paper.  This shows the cumulative probability that one,
two, or more publications listed as distinct to by both systems in
fact refer to the same publication.  For example, it shows that more
than 60\% of the articles are represented as unique entries within
\ISI's listing while 85\% of them are unique with \GS.  None of the
articles had more than five separate listings within \GS, while 13\%
had five or more entries in \ISI's system (e.g., the example shown in
Table 1 had 12).

Overlap between the two sources of data was relatively small.  Of the
203 citations analyzed, only 78 publications received at least one
cited reference from each system.  However, for this subset the
general pattern of agreement was quite good.  Figure
\pdfig{Correlation of \GS and \ISI citation counts}{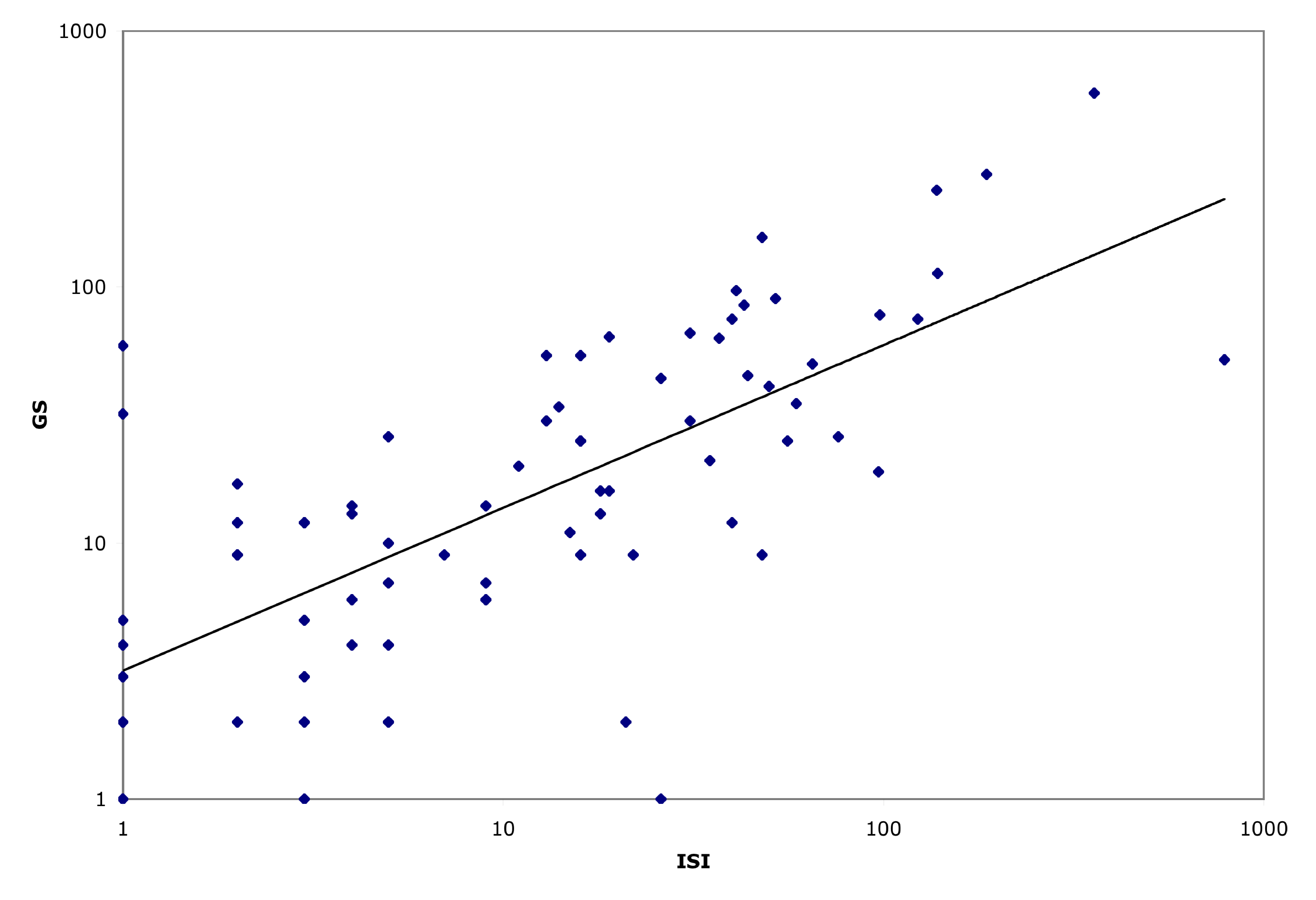}
shows the number of citations reported by \GS and \ISI for the subset
of 78 publications.  Note that the number of citations is plotted on a
log-log scale, reflecting the well-known power law distribution of
citation reference \cite{redner98}. Based on this sample, there seems
good evidence ($r^2 = 0.5023, t=8.872, \rho>0.005$) for a power law
relation ($GS = 3.1718 * ISI^{0.6359}$) relating the number of
citations reported by the two services.

Figure \pdfig{Temporal distribution of citations}{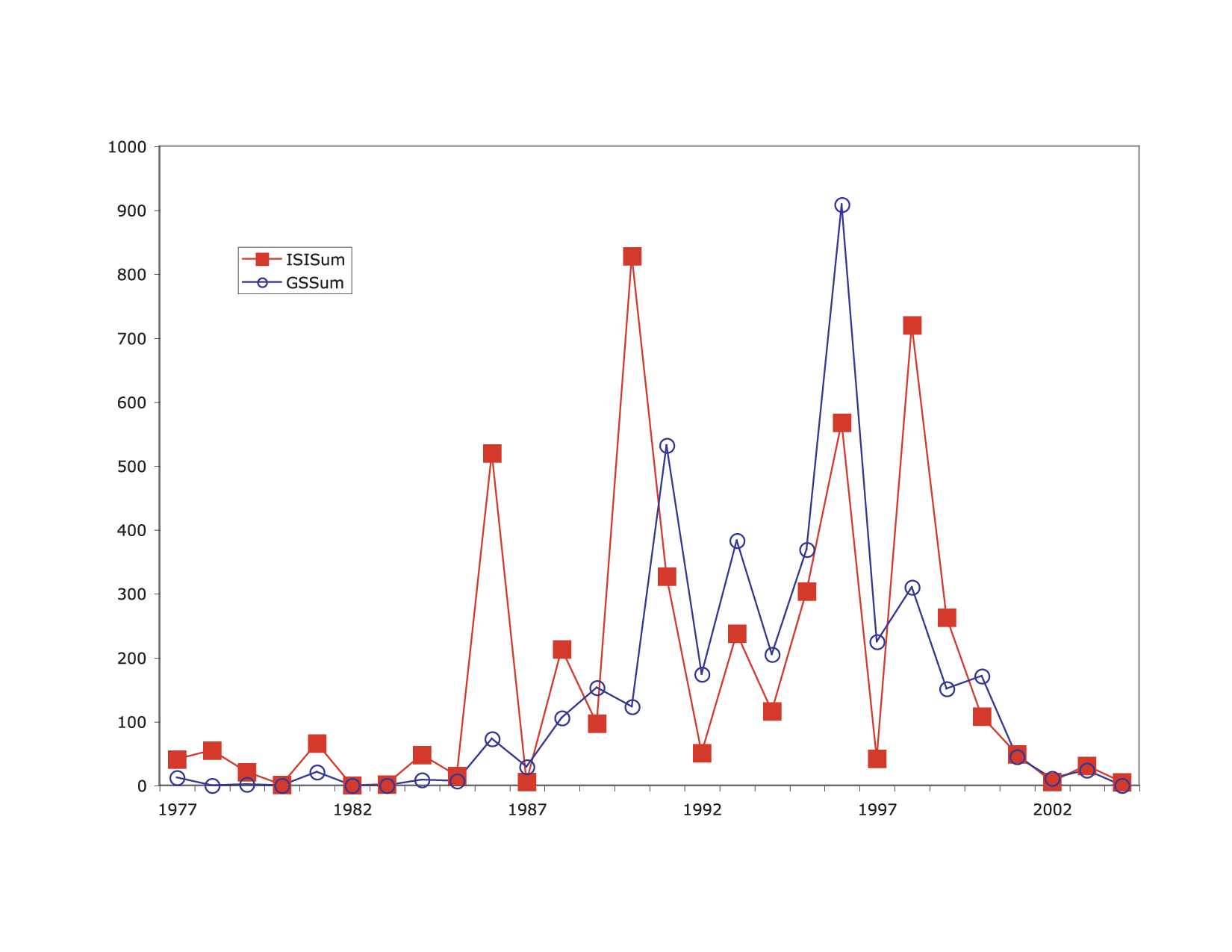} shows the
cumulative number of citations reported by publication year of the
cited work.  An alternative criterion for considering the match
between systems is to define a ``miss'' to be a publication for which
one service has identified three or more citations, but which the
other service does not capture whatsoever.  Figure \pdfig{Temporal
  distribution of missing citations}{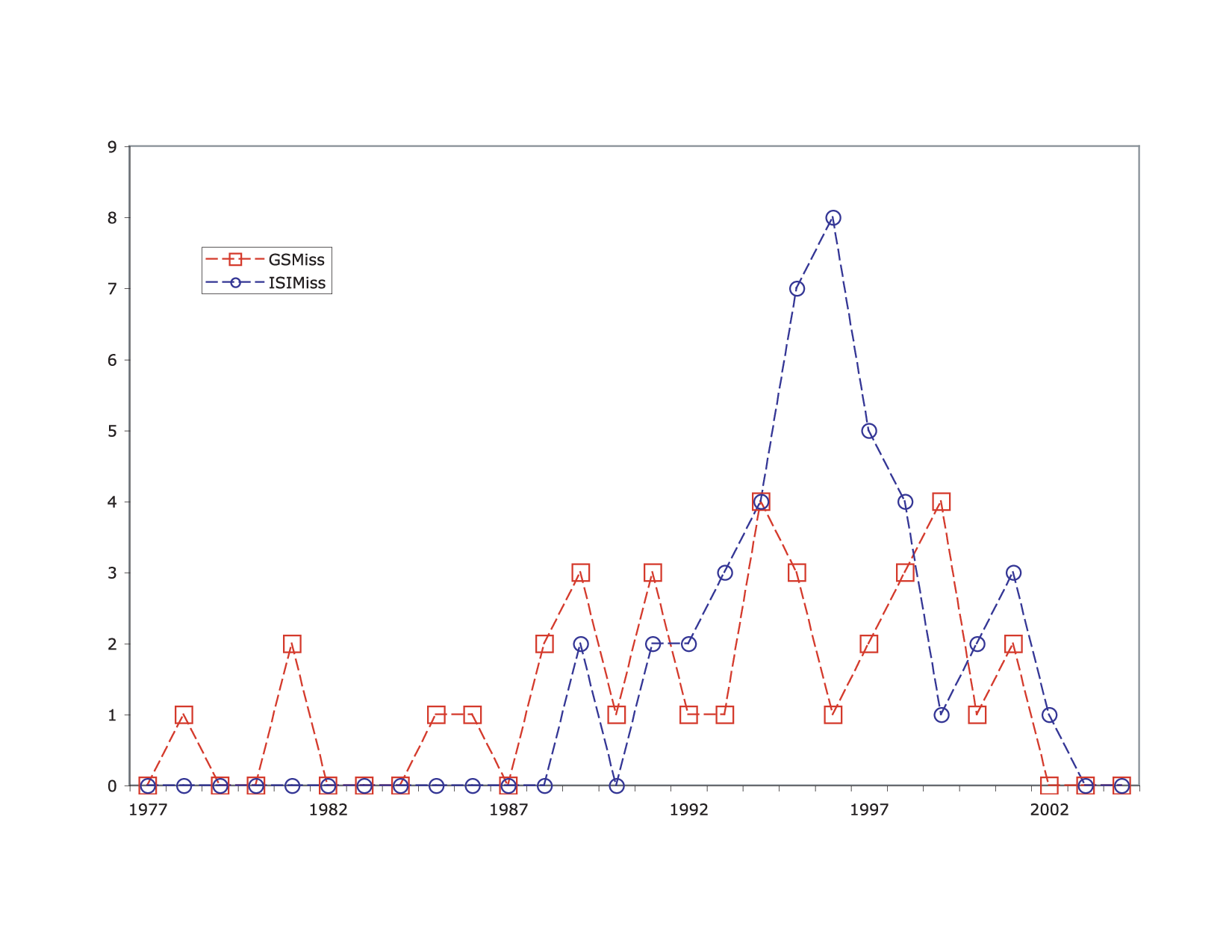} shows missing
citations, found by one service but not the other, again distributed
by publication year.  \GS seems competitive in terms of coverage for
materials published in the last twenty years; before then \ISI seems
to dominate.

Coverage with respect to the two systems can also be analyzed by other
dimensions of the publications, including publication venue and
author.  Figure \pdfig{Coverage by publication type}{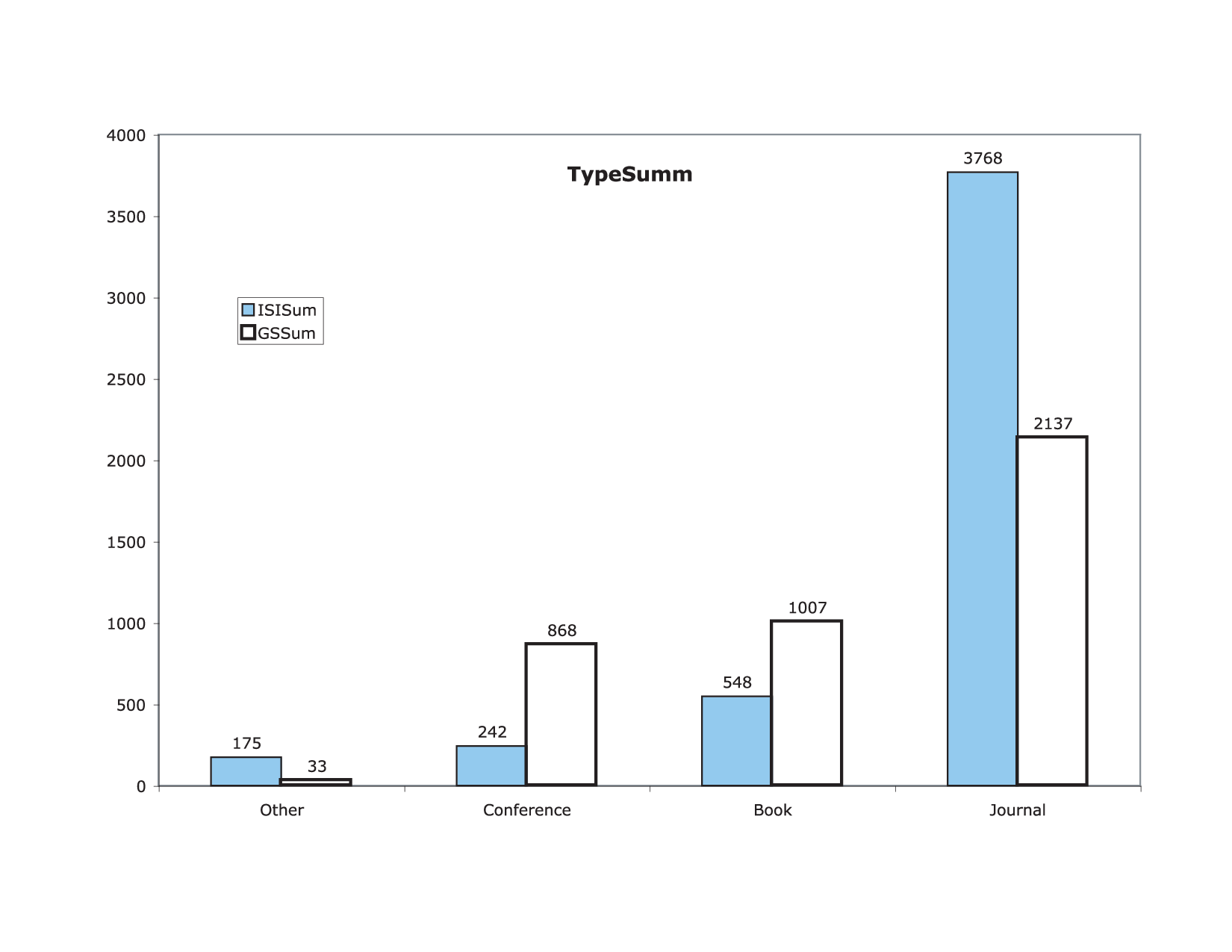}
aggregates publications into four categories: conference publications,
books (or book chapters), journal articles, and other forms of
publications (e.g, technical reports, dissertations, etc.); $\chi^2$
tests confirm the distributions are distinct.  Publications in books
(as noted by May, above) and conference proceedings are much more
likely to be available via \GS; conversely, journal articles are
better indexed via \ISI.  If citations are summarized with respect to
the six authors analyzed, Figure \pdfig{Coverage for individual
  authors}{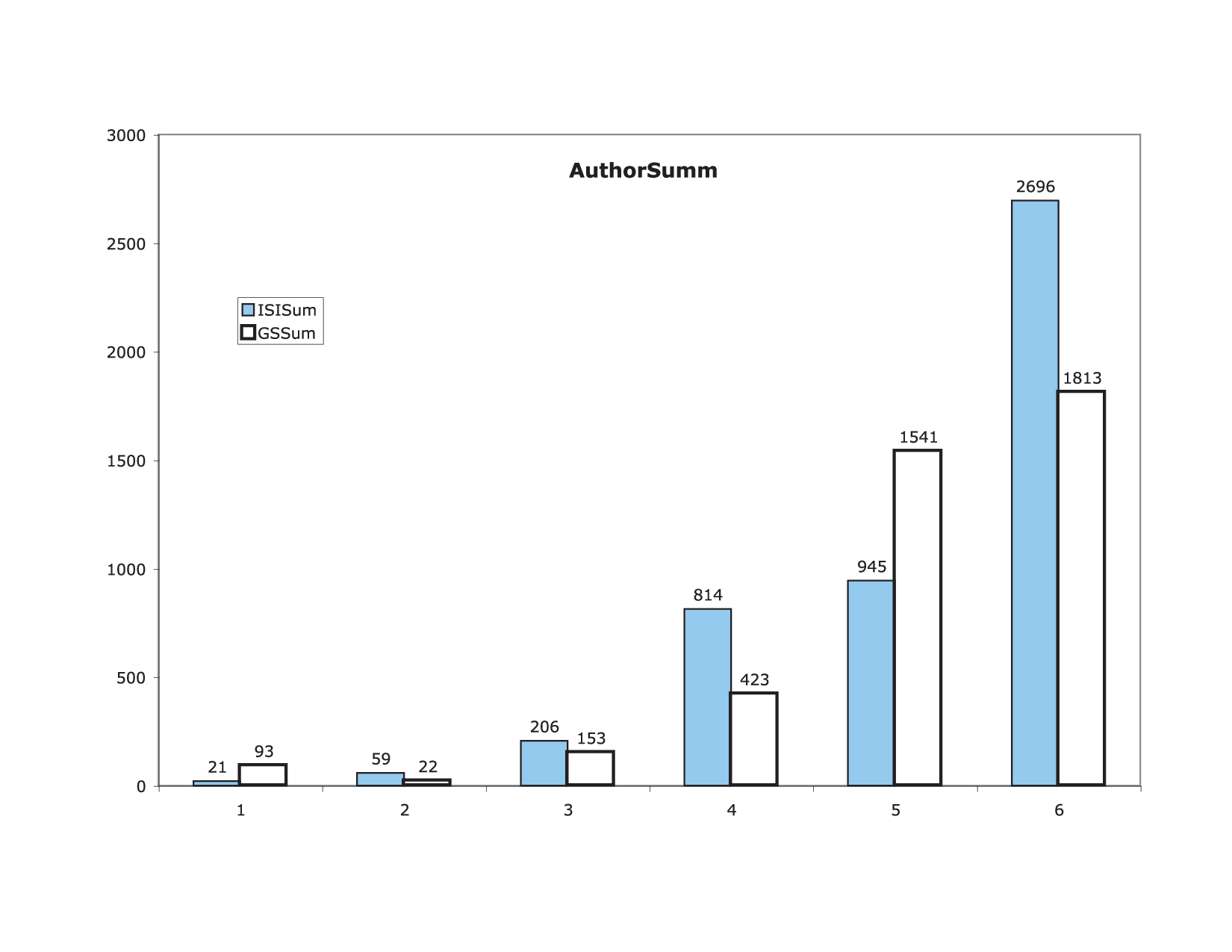} shows that some authors are better represented
with respect one service as opposed to another.  Such variation is to be
expected, given that some authors, via the publication venues through
which they typically report, will be more or less well-covered by one
service or another.  Again,  $\chi^2$
tests confirm the distributions are distinct.  

\section{Summary}

Evaluating academics' performance, as individuals or as part of larger
social groups, in terms of the number of publications they produce is
common practice.  The ability to quantify their ``impact'' in terms of
the number of other publications that subsequently choose to cite
their work arguably provides a more refined and relevant measure.
Such data is subject, however, to confounding factors ranging from
noise in the process of collating and ``inverting'' bibliographic
references through intrinsic features of scientific publication (e.g.,
self-citation).  The results presented above are therefore reassuring
in that new evidence provided by \GS provides the first
independent confirmation of impact data previously available only from
\ISI.  However, analysis across both systems also shows significant
variations with respect to the two dimensions (authorship and
publication type) considered; other dimensions of variation are
certain to exist.  This analysis also revealed some problems common to
both systems.  For example, both services support only simple ASCII
encodings of author names which are likely to lose important character
markup (available via Unicode representations) which can be especially
problematic for authors with foreign names.

Critically, new services within selected disciplines \cite{acmPortal,ieeeDL},
changing standards regarding exchange of ``open citation'' information
\cite{CrossRef}, in combination with increased pressure for public access to
scientific publications \cite{Zerhouni04}, may soon make some
operational difficulties associated with impact analysis obsolete. 
In the interim, academic deans,
science policy advisors and anyone else relying on citation count
data  are cautioned that any
individual measurement requires more context.  
In the longer term, the increased
availability of statistics like bibliographic impact makes it
increasingly important to understand how publication and citation
activities, within both scientific publication and Web publishing more
generally, can be included as part of more holistic evaluations of
intellectual contribution \cite{grant00}.

\bibliographystyle{plain}
\bibliography{iq,/Users/rik/Writing/FOA/Manuscript/foa.bib,/Users/rik/Writing/Biblio/belew.bib}

\end{document}